\documentclass[12pt]{iopart}

\usepackage{iopams}  
\usepackage{graphicx}  
\usepackage{float}
\usepackage{url}
\newcommand{\argmin}{\mathop{\mathrm{arg\,min}}}
\newcommand{\maximise}{\mathop{\mathrm{maximise}}}

\begin{document}

\title[Optimised surface-electrode ion-trap junctions]{Optimised surface-electrode ion-trap junctions for experiments with cold molecular ions}


\author{A. Mokhberi$^{1}$, R. Schmied$^{2}$, and S. Willitsch$^{1}$}
\address{$^{1}$Department of Chemistry, University of Basel, Klingelbergstrasse 80, 4056 Basel, Switzerland\\
$^{2}$Department of Physics, University of Basel, Klingelbergstrasse 82, 4056 Basel, Switzerland}
\date{\today}
\ead{\mailto {stefan.willitsch@unibas.ch}}
\begin{abstract}
We discuss the design and optimisation of two types of junctions between surface-electrode radiofrequency ion-trap arrays that enable the integration of experiments with sympathetically cooled molecular ions on a monolithic chip device. 
A detailed description of a multi-objective optimisation procedure applicable to an arbitrary planar junction is presented, and the results for a cross junction between four quadrupoles as well as a quadrupole-to-octupole junction are discussed. Based on these optimised functional elements, we propose a multi-functional ion-trap chip for experiments with translationally cold molecular ions at temperatures in the millikelvin range. This study opens the door to extending complex chip-based trapping techniques to Coulomb-crystallised molecular ions with potential applications in mass spectrometry, spectroscopy, controlled chemistry and quantum technology. 
\end{abstract}


\maketitle

\section{Introduction}\label{sec:Int}
Two-dimensional ion-trap arrays have become important tools for the development of large-scale quantum information processing \cite{wineland98a, kielpinski02a, home09a} and quantum simulation \cite{schmied09a, sterling13a, kumph16a} based on atomic ions.
In particular, surface-electrode (SE) radiofrequency (RF) ion traps in which all electrodes lie in a plane \cite{chiaverini05a} offer the possibility of integrating multiple trapping zones and of precisely shaping electrodes, and therefore, trapping potentials. The scalability of SE structures offered by microfabrication methods paves the way for novel large-scale experiments \cite{hughes11a}.

Translationally cold and quantum-state controlled molecular ions, on the other hand, are promising platforms for fundamental studies such as the test of the time invariance of physical constants \cite{biesheuvel16a}, as well as for applications ranging from precision spectroscopy \cite{loh13a, germann14a, asvany15a,koelemeij07b} and metrology \cite{schiller14a} to collision studies \cite{hall12a} and controlled chemistry \cite{ospelkaus10b, chang13a}. The workhorse technique for preparing such systems is sympathetic cooling with laser-cooled co-trapped atomic ions \cite{willitsch12a} in RF traps \cite{major05a}. This method enables the formation of Coulomb-crystallised molecular ions at temperatures in the millikelvin range\cite{willitsch12a}. Recently, the sympathetic cooling of molecular ions in a SE RF trap has been demonstrated \cite{mokhberi14a}. 
The extension of SE trapping technology to cold molecular ions opens up exciting possibilities, e.g., the implementation of miniaturised guided-ion-beam experiments which could combine mass spectrometry, spectroscopy, reaction studies, and cold chemistry experiments on a chip device.

In order to develop an ion-trapping network, the key ingredient is junctions. Usually, a junction is thought of as a connection between identical trap arrays, e.g., a T-, Y-, or X- (cross) junction connecting quadrupolar channels \cite{hensinger06a, blakestad09a, amini10a, wright13a}. In addition, a junction can be seen as an element that enables the modification of trapping fields between two different trapping configurations, e.g., a quadrupole-to-octupole junction which can be of interest for, e.g., connecting a standard quadrupole ion transport channel with a large-volume octupolar ion storage region.  
Such field-modifying junctions allow the manipulation of the ion-crystal structures as well as the separation of a single harmonic well into a double-well trapping configuration with potential applications in quantum information and quantum matter-wave experiments \cite{hammer15a}. 
 
The main challenge to the use of SE junctions arises from the fact that the lowest-order multipole component of the RF field at the centre of a given intersection is a hexapole determined by the tangents of the intersecting channels \cite{wesenberg09a}. In RF traps, the trapping potential is usually calculated within the adiabatic approximation \cite{gerlich92a} in which a time-independent pseudopotential derived from the RF field governs the confinement of ions \cite{major05a}. Straight RF electrodes intersecting at right angles do not provide pseudopotential confinement perpendicular to the paths due to field cancellation. The pseudopotential at the centre becomes stiffer as the angle between straight arms is reduced and also as the number of intersecting channels decreases. For this reason, SE Y-junctions \cite{amini10a, moehring11a} were initially preferred to SE X-junctions \cite{wright13a}. 
 
X-junctions are particularly important for applications which need to economize space, e.g., highly integrated experiments. Recently, efforts have been invested in developing such junctions resulting in their successful demonstration and in new techniques for better controlling ion shuttling operations \cite{wright13a, satellite13a}. 
One of the key elements for these achievements is the optimisation of the RF-electrode geometry in the vicinity of the intersections such that ions are confined along the direction perpendicular to the surface \cite{wright13a}.

The ion-trap junctions that have been developed thus far were designed to enable the shuttling of single-species atomic ion crystals while minimising ion heating during transport \cite{amini10a, moehring11a,wright13a,shu14a}.
Besides the applications in molecular physics mentioned above, the extension of this technology to bicomponent Coulomb crystals is also important for quantum technology and quantum computation with mixed-species crystals \cite{home16a}. The challenge for the transportation of bicomponent crystals through junctions arises from the mass dependence of the pseudopotential.

In this paper, we describe a general method for the optimisation of an arbitrary planar junction in accordance with the requirements for the transportation of mixed-species ion crystals (section \ref{sec:MOO}). This approach is applied to design a cross between four quadrupoles and to a quadrupole-to-octupole junction discussed in detail in section \ref{sec:Res_Ana}.  
Based on this study, a design of a multi-functional ion-trap array which enables the integration of several experiments on a single-layer chip is presented (section~\ref{sec:MFIT}).  

\section{Multi-objective optimisation of electrode structures}\label{sec:MOO}

\subsection{Objective functions and weighted sum method}\label{subsec:objfunc}
Transporting ions through SE junctions requires the minimisation of several objectives.
A variety of objective functions has been considered \cite{amini10a, moehring11a, wright13a, liu14a}, and thus, the question arises which physical quantities are relevant when optimising an electrode structure for a given application. 
First and foremost, the suppression of pseudopotential barriers that impede ion transport is crucial \cite{wesenberg09a}. Note that the position of pseudopotential minima, where the RF field vanishes, solely depends on the trap geometry, and thus is identical for all ion masses. However, the mass dependence of the pseudopotential may cause a segregation of laser-cooled and sympathetically cooled ions during the transportation over barriers which could lead to insufficient cooling and high loss rates of molecular ions. Thus, for shuttling mixed-species ion crystals the height of pseudopotential barriers is the most important quantity to be minimised (objective 1).

For reliable shuttling of ion qubits, axial pseudopotential gradients that cause ion heating must be suppressed \cite{amini10a, moehring11a, wright13a, shu14a, satellite13a}, since at intersections, where noise on the RF potential is present, axial motional modes of ions might be excited \cite{blakestad09a}. Thus, for applications in quantum computation the minimisation of the pseudopotential gradient is crucial (objective 2). 
An alternative approach to near-motional-ground-state shuttling of ions is the minimisation of the variation of the axial potential curvature \cite{satellite13a}. 
We note that this optimisation objective is less important in the present context as the molecular ions are continuously sympathetically cooled by laser-cooled ions, also during shuttling.

Furthermore, the variation of the trapping height needs to be controlled for two reasons. First, weakly confining potentials at SE intersections might cause significant ion losses. This issue can be mitigated by using static fields that forces ions to circumnavigate the junction centre at the expense of moving the trapping minima away from the pseudopotential null \cite{wright13a}. 
This technique is unsuited for bicomponent crystals due to the risk of the drastic segregation of ionic species under the influence of such a static field \cite{mokhberi15a}. Second, for the transportation of molecular ions together with laser cooled atomic ions in the presence of cooling beams large variations of the trapping height should be avoided (objective 3). 

These three objectives evaluated at the ion channel ${\mathbf{r}_{\rm min}(x)=\argmin \limits_{y,z}({\Phi}_{\rm ps}(x,y,z))}$, where $\argmin$ represents the values of \emph{y} and \emph{z} that minimise ${\Phi}_{\rm ps}(x,y,z)$ at a certain \emph{x}, can be cast into the following mathematical form:
\numparts
\begin{eqnarray}
\label{eqn: obj functions1}
F_{1}&=\int_{-L}^{0}{\mathit{\Phi}_{\rm ps}(\mathbf{r}_{\rm min}(x))dx},\\
\label{eqn: obj functions2}
F_{2}&=\int_{-L}^{0}{\parallel \frac{\partial \mathit{\Phi}_{\rm ps}}{\partial x} (\mathbf{r}_{\rm min}(x))\parallel^2 dx},\\
\label{eqn: obj functions3}
F_{3}&=\int_{-L}^{0}{(z_{\rm min}-h)^2 dx}.
\end{eqnarray}
\endnumparts
$F_{i}$ ($i$=1, 2, 3) denotes the $i$th objective function, $\mathit{\Phi}_{\rm ps}$ is the pseudopotential, $L$ is the coordinate of the initial trapping zone, $z_{\rm min}$ is the component of ${\mathbf{r}_{\rm min}(x)}$ perpendicular to the chip surface, and $h$ is the intended trapping height. 
Note that the minimisation of pseudopotential barriers might occur at the cost of large trapping height variations and vice versa.

An efficient method of solving multi-objective optimisation problems is the weighted sum method in which a function $U$ is minimised \cite{marler05a}: 
\begin{eqnarray}\label{eqn:obj_total}
U&=\sum^{3}_{i=1} w_{i} [\xi_{i}F_{i}(d_{1}, ...,d_{p})]\\ \nonumber
&=\lbrace w_{1}\xi_{1}, w_{2}\xi_{2}, w_{3}\xi_{3}\rbrace \cdot \lbrace F_{1},F_{2},F_{3}\rbrace .
\end{eqnarray} 
Here, $w_{i}>0$ ($i$=1, 2, 3) is the weighting factor for the $i$th objective function $F_{i}$ which determines a comparative contribution specifing the user's preference. The lower-bound transformation method \cite{marler05a} was used to ensure \emph{F$_{i}$} are of comparable magnitudes in numerical calculations. Thereby, the transformation coefficient $\xi_{i}$ ($i$=1, 2, 3) was given by: 
\begin{equation} \label{eqn:transformation}
\xi_{i}=\frac{1}{F_{i}^{0}} ,
\end{equation}
where $F_{i}^{0}$~=~minimum $\lbrace F_{i}(d_{1}, ...,d_{p})\rbrace$, and $d_{1-p}$ denote the $p$ degrees of freedom, see section~\ref{subsec:design space}. 
As highlighted by equation~(\ref{eqn:obj_total}), it is important to notice a subtle distinction between using weights to determine the user's preferences and using weights to transform objective functions. The $w_{i}$ are arbitrary values systematically varied in the interval (0,1) to set the relative importance of the objectives; see, e.g., the specified values of $w_{i}$ and $\xi_{i}$ presented in appendix A.

In our calculations, we used normalized (non-dimensional) objective functions $F_{i}$ in equation (1). A non-dimensional pseudopotential can be written as $\widetilde{\mathit{\Phi}}_{\rm ps}={{\mathit{\Phi}}_{\rm ps}}({\frac{\ q^{2} V_{\rm RF}^{2}}{4m \Omega_{\rm RF}^{2}h^{2}}})^{-1}={\Vert \nabla \mathit{\Theta}_{\rm RF}\Vert}^2$, where $\mathit{\Theta}_{\rm RF}$ is the RF-electrodes basis function \cite{hucul08a} which depends solely on the geometry. Here, $m$ and $q$ are the mass and charge of the trapped ions, $h$ is the trapping height, and $V_{\rm RF}$ and $\Omega_{\rm RF}$ denote the RF drive amplitude and angular frequency, respectively.
Building on this formalism, the objective functions and, hence, solutions of the optimisation are independent of the mass of the ions, the trapping parameters ($V_{\rm RF}$ and $\Omega_{\rm RF}$)  as well as the trapping height. Note that owing to the normalisation of $F_{3}$ in equation (\ref{eqn: obj functions3}) the trapping height, which is a characteristic length of the system, does not contribute to our optimisation calculations, and can be substituted in the final result. As a result, the geometries obtained (see appendix A) present universal SE structures applicable to any mass or set of trapping parameters. 

\subsection{Design space}\label{subsec:design space}
\begin{figure}
\includegraphics[scale=0.38]{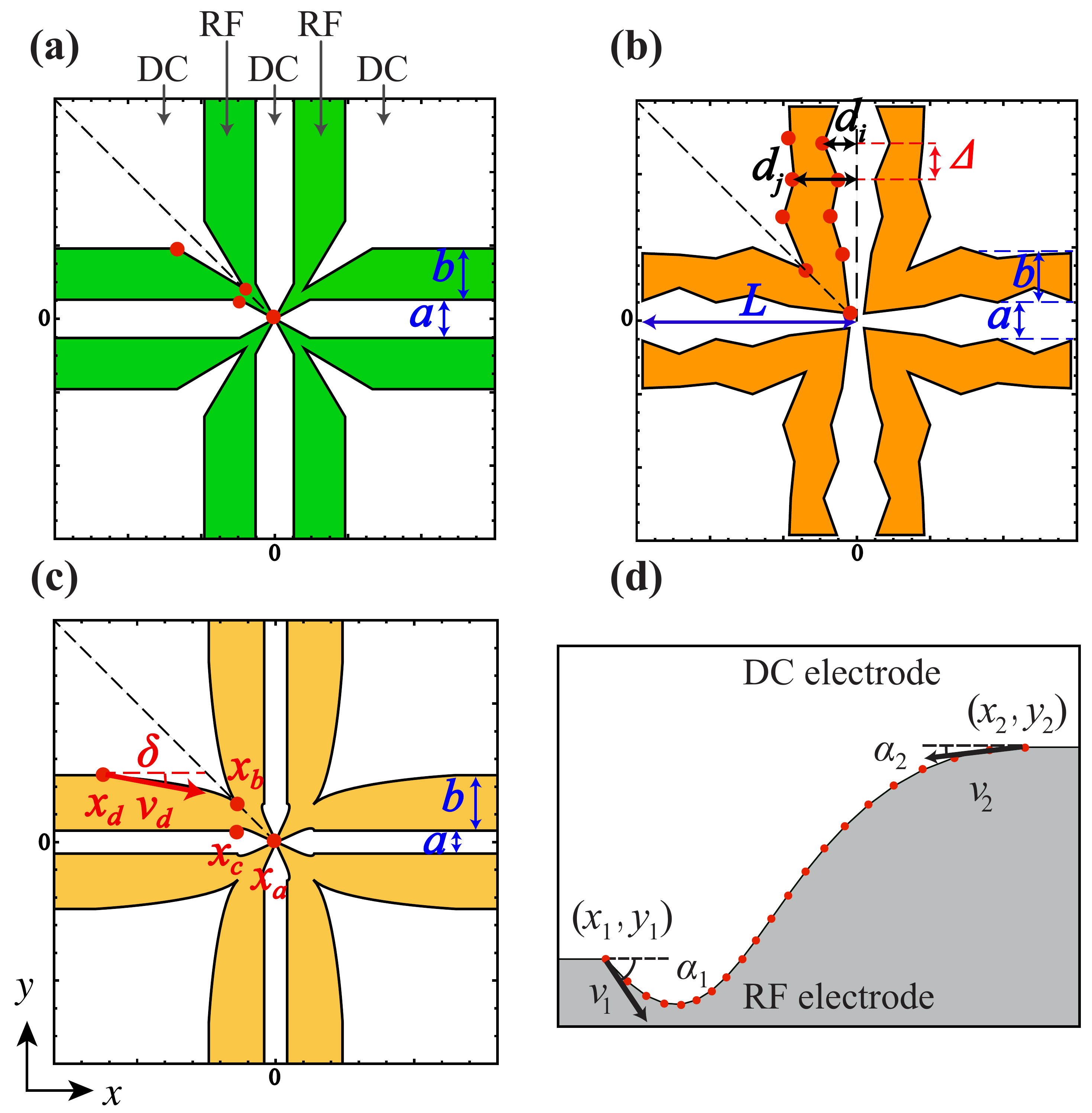}
\centering
\caption{\label{fig:design-space} Parametrisation of the cross junction geometry for the first optimisation. In (a-c), the RF electrodes are shown in colour, the red points indicate the control points, and $a$ and $b$ are the width of central DC and RF electrodes, respectively. The control points are connected via lines in (a), line elements defined by $d_{i}$, their distance $\mathit{\Delta}$ and the involved channel length \textit{L} in (b), and cubic splines in (c)-(d). (d) shows the eight parameters specifying each cubic spline, i.e., $\lbrace x_{1}, y_{1}, v_{1}, \alpha_{1} \rbrace$, $\lbrace x_{2}, y_{2}, v_{2}, \alpha_{2} \rbrace$, which are the positions of the initial and final points and the amplitude as well as the orientation angle of the first derivatives of the electrode boundary function at these points. In (c), each point, labelled with its abscissa value, exhibits three degrees of freedom, and the electrode geometry was fully defined by $\lbrace -x_{d}, a/2+b, v_{d}, \delta \rbrace$, $\lbrace -x_{c}, a/2, v_{c}, \gamma \rbrace$, $\lbrace -x_{a}, x_{a}, v_{a}, \alpha \rbrace$, and $\lbrace -x_{b}, x_{b}, v_{b}, \beta \rbrace$. Note that only one eighth of the geometry was parametrised due to dihedral D$_{4}$ symmetry in every cross junction as highlighted by dashed lines.}
\end{figure}

In this study, the design space refers to a set of parameters defining RF electrode geometries to be used in optimisation processes. 
The number and locations of control points specifying RF electrode shapes were varied according to the three methods shown with the example of a cross junction in figure~\ref{fig:design-space}. 
The first approach utilized four control points, where two points on each perimeter are connected simply via a line (figure~\ref{fig:design-space}(a)). The choice of only four parameters accelerated optimisation processes and led to satisfactory solutions despite the low number of degrees of freedom.

In the second approach, the inner and outer RF electrode edges were broken up into several line elements (figure~\ref{fig:design-space}(b)). 
We began with four control points and increased them up to 20 points while aiming for improved results. Careful attention must be paid when increasing the number of degrees of freedom because it does not necessarily result in better solutions due to the higher probability of getting trapped in local minima of the optimisation function.
This method of parametrisation has been widely used \cite{amini10a, moehring11a, wright13a, satellite13a, hammer15a, liu14a}, however, electrode structures obtained often exhibited sharp features arising from underdetermined geometrical constraints. Fourier components of the surface potential of wavelength $\lambda$ exponentially decrease as exp($-\frac{2\pi z}{\lambda}$) \cite{wesenberg08a}, and consequently, the field is not significantly influenced by geometric elements of size $\lambda\ll z$. Such sharp features \cite{moehring11a, wright13a} might also hinder precise manufacturing and would reduce breakdown voltages through the bulk or surface flashover.

Thus, we employed a third approach to mitigate these issues by using cubic splines to define the design. Control points on each electrode perimeter were bound to a cubic spline specified by eight parameters, see figure~\ref{fig:design-space}(c)-(d). Because of symmetry, the number of parameters for each spline was eventually reduced to six, and thus the geometry of the cross junction was defined by 12 parameters. This method of parametrisation was found to be flexible as well as efficient and was mainly used in our calculations. 

\subsection{Implementation}\label{subsec:algorithms}
In our calculations, a two-step optimisation process was applied. In the first optimisation, the potential generated by SE electrodes was analytically calculated in the ``gapless plane approximation'' \cite{schmied10a} using the \textit{SurfacePattern} software package \cite{SurfacePattern12}. These fast calculations enabled many iterations, and hence the exploration of some of the generic aspects of junction geometries. Due to weakly confining potentials at ion channel intersections, the effect of gaps cannot be neglected, and therefore, re-optimisation of the structure in the presence of the gaps is essential. The geometries obtained from the first optimisation served as starting points for the second, in which electric potentials generated by three-dimensional structures including gaps were computed using the finite element method (FEM) \cite{COMSOL}. 
In a multi-objective optimisation process, a single set of parameters that simultaneously minimises all objective functions usually does not exist, hence, a \textit{Pareto optimal} solution is searched among many possible solutions \cite{marler05a}.

\section{Optimisation results and analyses}\label{sec:Res_Ana}
\subsection{Optimisation of the cross junction}\label{subsec:cross junction first}
\begin{table}
\caption{\label{table:opti_quadchanel}Optimal electrode configurations for a typical five-wire surface-electrode trap as the asymptotic quadrupolar channel for the cross junction. $a$ and $b$ are the RF electrode separation and width, respectively, $h$ is the trapping height above the surface. See text for discussion. }
\renewcommand{\arraystretch}{1.5}
\begin{indented}
\item[]\begin{tabular}{@{}l l l p{1.5cm} p{1.5cm} p{4.55cm}}
\hline
\mr
Reference&\textit{a} &$b/a$ &Relative trap depth at fixed $h$&Relative channel curvature at fixed $h$& Method \\
\mr
\cite{house08a}&1.09~$h$&\textbf{1.19}&68\%&83\%&maximum depth at fixed $a$\\
\cite{wesenberg08a, nizamani11a}&0.69~$h$&\textbf{3.68}&100\%&94\%&maximum depth at fixed $h$\\
\cite{schmied09a, wesenberg08a}, this work&0.83~$h$&\textbf{2.41}&95\%&100\%&maximum curvature at fixed $h$\\
\mr
\hline
\end{tabular}
\end{indented}
\end{table}
First, it is worth discussing the geometries of intersecting RF rails for infinite quadrupolar channels.  
In reference \cite{house08a}, the optimal ratio of the width of the RF electrodes $b$ to their separation $a$ in a five-electrode design (DC-RF-DC-RF-DC) with  RF electrodes of equal width (see figure \ref{fig:design-space}) was calculated for a given $a$ while maximising the trapping depth in the absence of control static fields (referred to as the intrinsic depth, that is, the value of the pseudopotential at escape points). 
Because of the importance of the trapping height $h$ as a constraint in many experiments, the intrinsic depth as well as the electric potential curvature was optimised for a fixed $h$ in references \cite{schmied09a, wesenberg08a,nizamani11a}. 
Table \ref{table:opti_quadchanel} presents a detailed comparison between these three calculations, highlighting the fact that the optimal value of $\frac{b}{a}$ depends on the optimisation goal.

\begin{figure}[h]
\centering
\includegraphics[scale=1.1]{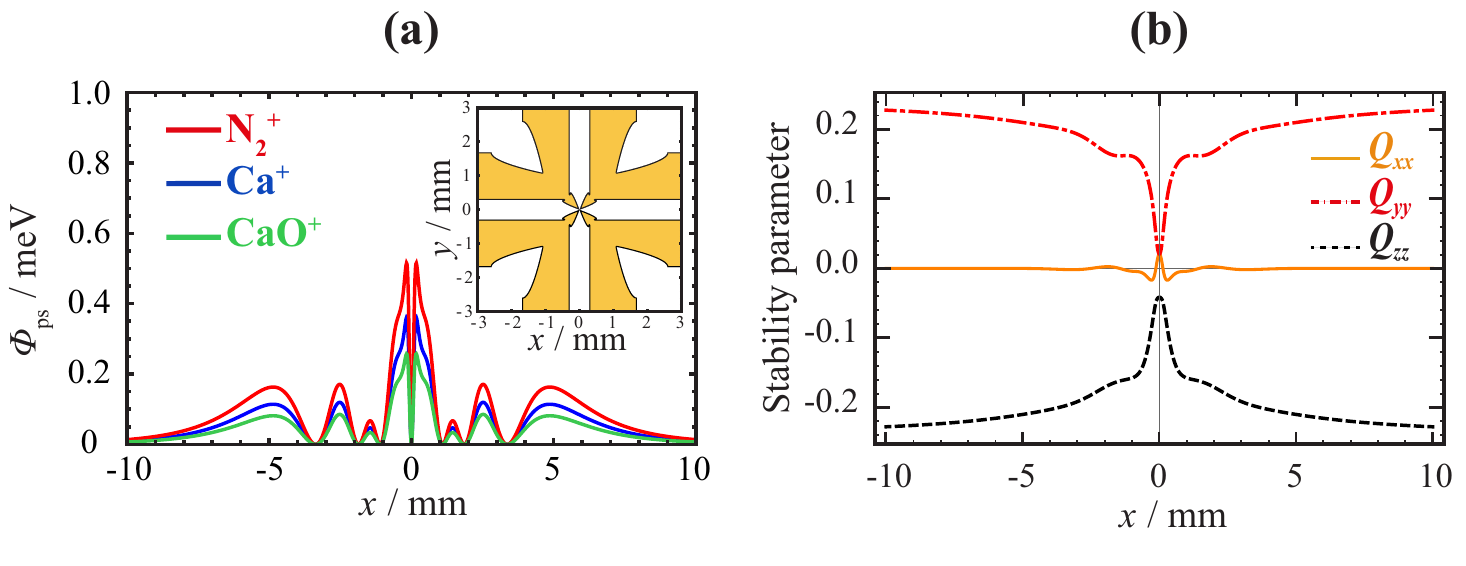}
\caption{\label{fig:crossjunction-realsize}The result of the first optimisation of the cross junction. (a) Pseudopotential barriers for N$_{2}^{+}$, Ca$^{+}$, and CaO$^{+}$. The inset depicts the corresponding optimal geometry. (b) Generalised stability parameters \cite{house08a} calculated for Ca$^{+}$ along the ion channel. }
\end{figure}

We fixed the ratio $\frac{b}{a}$ to an optimised value for a specific pseudopotential curvature in the transverse directions, and used the geometric degrees of freedom in the vicinity of the junction to suppress barriers. 
We chose the value $\frac{b}{a}=2.414$, which maximises the curvature of the RF field in the transverse plane and additionally leads to the trapping depth of 95\% of the maximum achievable one at a given trapping height (table~\ref{table:opti_quadchanel}). 

We chose the trapping height to be $\approx$~700~$\mu$m, and hence, the width of the RF electrodes $b=$~1400~$\mu$m and their separation $a=$~580~$\mu$m. Thus, a sufficiently large trapping volume for a few hundred ions as well as stable trapping conditions for a variety of interesting ion species can be achieved (see below and figure \ref{fig:crossjunction-realsize}). Moreover, a suitable choice of the trap size is crucial to eliminate effects of anharmonic terms in the trapping potential which increase with decreasing trap dimensions. These effects are pronounced in SE traps owing to their asymmetric geometries, and their consequences on bicomponent Coulomb crystals, e.g., the segregation of atomic from molecular ions, have been discussed in \cite{mokhberi15a}.

The minimisation of pseudopotential barriers is assessed based on a suppression factor which is defined as the ratio between the height of the suppressed barriers and either those of a straight junction or the intrinsic depth of a straight channel far away from the junction. 
The result of the first optimisation of the cross junction is presented in figure~\ref{fig:crossjunction-realsize}. The normalised pseudopotential is evaluated for the RF amplitude V$_{\rm RF}$=300~V, and the RF angular frequency $\Omega_{\rm RF}$=$2\pi\times10.0$~MHz. In this case, the highest barrier for $^{40}$Ca$^{+}$ is 0.36~meV (figure~\ref{fig:crossjunction-realsize}(a)), 614 times less than the depth of a straight channel, and 125 times less than the barriers present in a design with completely straight electrodes. 

The non-confining characteristic of the trapping potential at the intersection region manifests itself in a drastic change of the stability parameters as illustrated in figure~\ref{fig:crossjunction-realsize}(b). 
The $Q_{ii}$ ($i\in\lbrace x,y,z\rbrace$) in figure~\ref{fig:crossjunction-realsize}(b) are the diagonal elements of the multi-dimensional form of the stability parameter $q_{\rm M}$ of the Mathieu equation \cite{shaikh12a}, which are defined by
\begin{equation} \label{eqn:transformation}
Q_{ii}(x)=\frac{2 q V_{\rm RF}}{m \Omega_{\rm RF}^{2}}\frac {\partial^{2}{\mathit{\Theta}_{\rm RF}}}{{\partial x_{i}}^{2}}(\mathbf{r}_{\rm min}(x)),
\end{equation} 
evaluated along the ion channel ${\mathbf{r}_{\rm min}(x)=\argmin \limits_{y,z}({\Phi}_{\rm ps}(x,y,z))}$, where $\argmin$ represents the values of \emph{y} and \emph{z} that minimise ${\Phi}_{\rm ps}(x,y,z)$ at a certain \emph{x}, $\mathit{\Theta}_{\rm RF}$ is the RF-electrodes basis function \cite{hucul08a}, and $m$ and $q$ are the mass and charge of the trapped ions, respectively.

\subsection{Second optimisation of the cross junction and characterisation of a central bridge}\label{subsec:cross junction second}
\begin{figure}[h]
\includegraphics[scale=0.35]{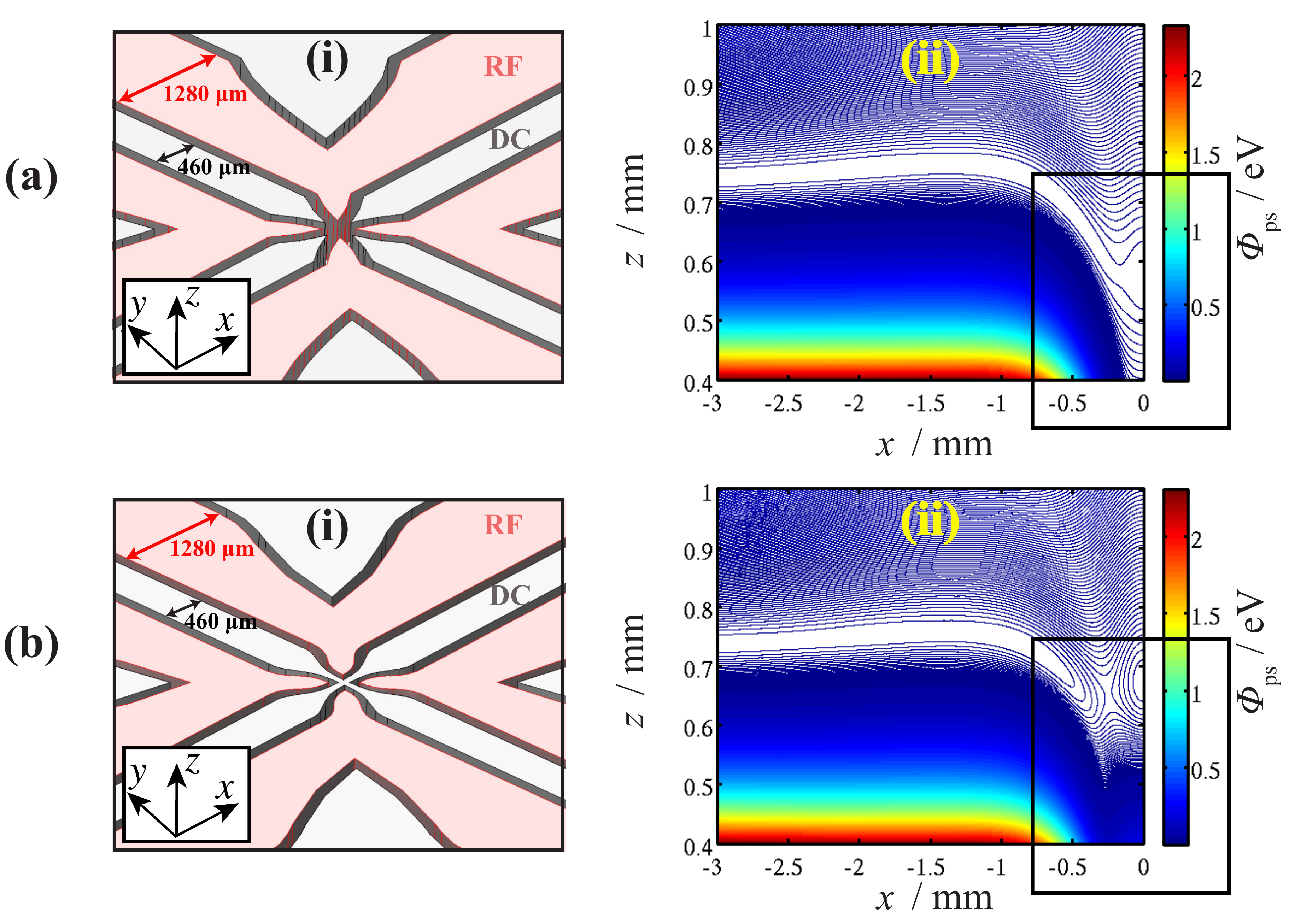}
\centering
\caption{\label{fig:Ps-contour}Results of the first (a) and second (b) optimisation of the cross junction. The latter features a 70-${\mu}$m-wide bridge between the central DC electrodes. The corresponding pseudopotentials calculated for $^{40}$Ca$^{+}$ in the (\textit{x,z}) plane labelled with (ii) show a loosely confining region at the junction centre in (a), whereas a stiff trapping potential is achieved in (b) (see the boxed areas). The isopotential contours are separated by 1~meV.}
\end{figure}
The three-dimensional structure of the cross junction with the inclusion of 120~$\mu$m wide gaps was modelled using FEM (figure~\ref{fig:Ps-contour}(a)~(i-ii)). Due to the 330~$\mu$m diameter hole at the centre, the pseudopotential exhibited a spatially broad, loosely confining trapping region along the vertical direction $z$ (figure~\ref{fig:Ps-contour}(a)~(ii)). 

\begin{figure}
\includegraphics[scale=0.55]{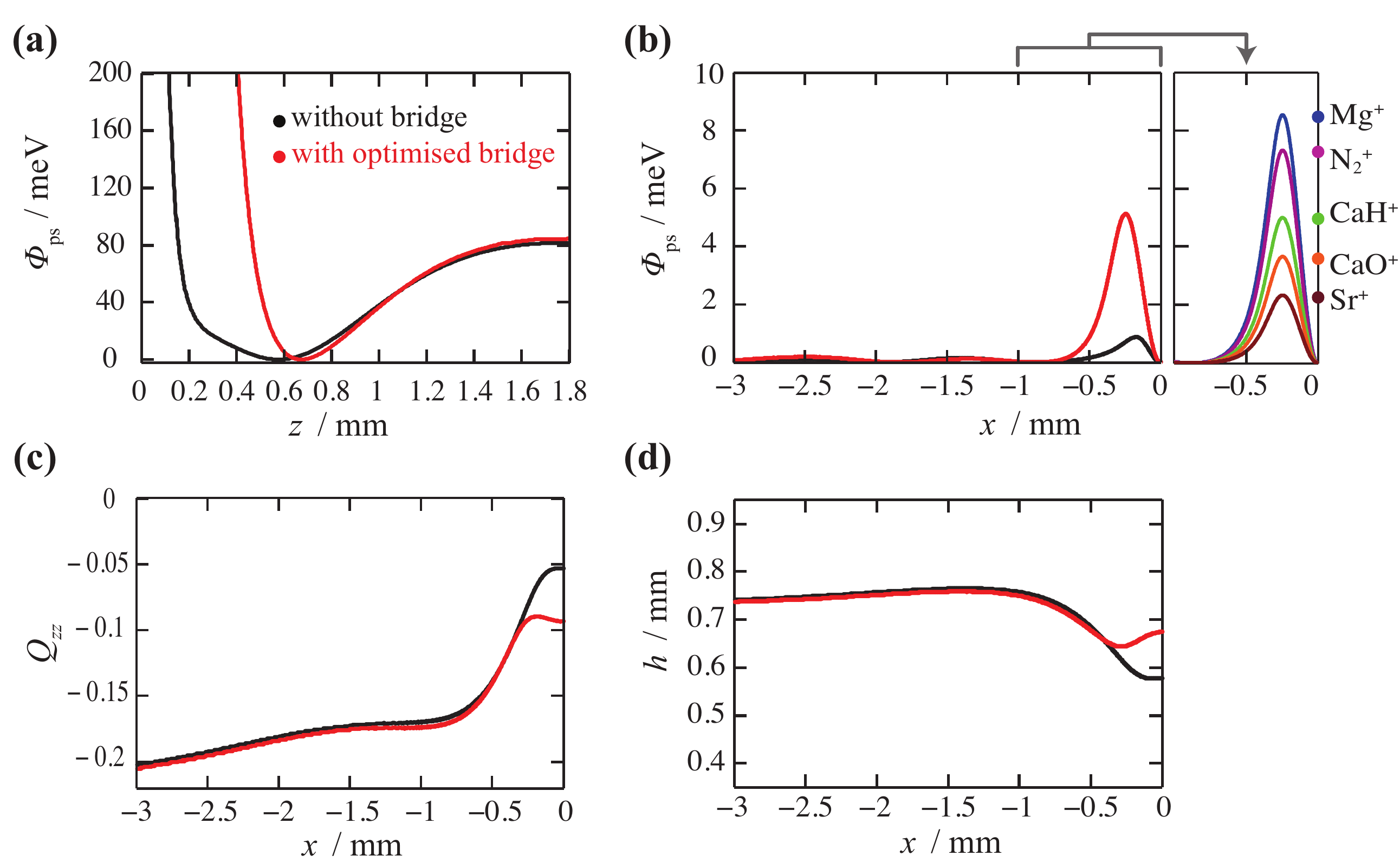}
\centering
\caption{\label{fig:conclusion-bridge}Comparison of the features of the two cross junctions depicted in figure~\ref{fig:Ps-contour}: without bridge (black) and  with optimised bridge (red). (a) One-dimensional cuts through pseudopotentials along the vertical axis \textit{z} at the centre of the junction. (b) The pseudopotential barriers, (c) the variation of $Q_{zz}$, and (d) the variation of the trapping height along ion channels. Note that all calculations are presented for $^{40}$Ca$^{+}$, whereas the inset in (b) displays the highest barriers for the structure with the bridge for different ion species.}
\end{figure}

To address this issue, we investigated the incorporation of a bridge between the central DC electrodes with a carefully refined geometry. To explore the consequences of this modification, over 100 geometries were simulated using FEM. Bridges formed by merging the tips of RF electrodes at the centre cause extremely high pseudopotential barries. 
This situation is opposed to junctions in three-dimensional electrode geomeries, e.g., the one demonstrated in \cite{blakestad09a}, where RF bridges were located symmetrically above and below the ion channel. 
Our calculations show that as the bridge widens, pseudopotential barriers drastically increase. Further calculations were performed using a 70-${\mu}$m wide DC bridge in order to enhance mechanical stability of the structure \cite{mokhberi16a}. 
For a fixed DC bridge width, the height of pseudopotential barriers reduces as the size of the gaps in the vicinity of the junction centre decreases, and thus a further decrease of the barriers depends on how narrow the gaps can be fabricated. 
These calculations enabled the suppression of pseudopotential barriers by a factor of three compared to an initial straight bridge of the same width while the variation of the trapping height was controlled in a range of $\pm$10\%. The resulting geometry depicted in figure~\ref{fig:Ps-contour}(b)~(i) features a confining trapping potential perpendicular to the ion channel at the intersection (figure~\ref{fig:Ps-contour}(b)~(ii)). 

Figure~\ref{fig:conclusion-bridge} illustrates the influence of the incorporation of the optimised DC bridge on the characteristics of the cross junction. A significant increase of the trap stiffness at the centre has been achieved (figure~\ref{fig:conclusion-bridge}(a)), which also manifests itself in the larger $Q_{zz}$ parameter improved by a factor of two (figure~\ref{fig:conclusion-bridge}(c)). This junction exhibits pseudopotential barriers~$<$~5.5~meV for $^{40}$Ca$^{+}$ (figure~\ref{fig:conclusion-bridge}(b)), which is less than 2.5\% of the trap depth far from the junction in the quadrupolar channel, whilst the ion channel is elevated at a well-controlled height over the ion shuttling path (figure~\ref{fig:conclusion-bridge}(d)). 

\subsection{Quadrupole-to-octupole junction}\label{sec:octu-to-quadru junction}
\begin{figure}
\includegraphics[scale=0.6]{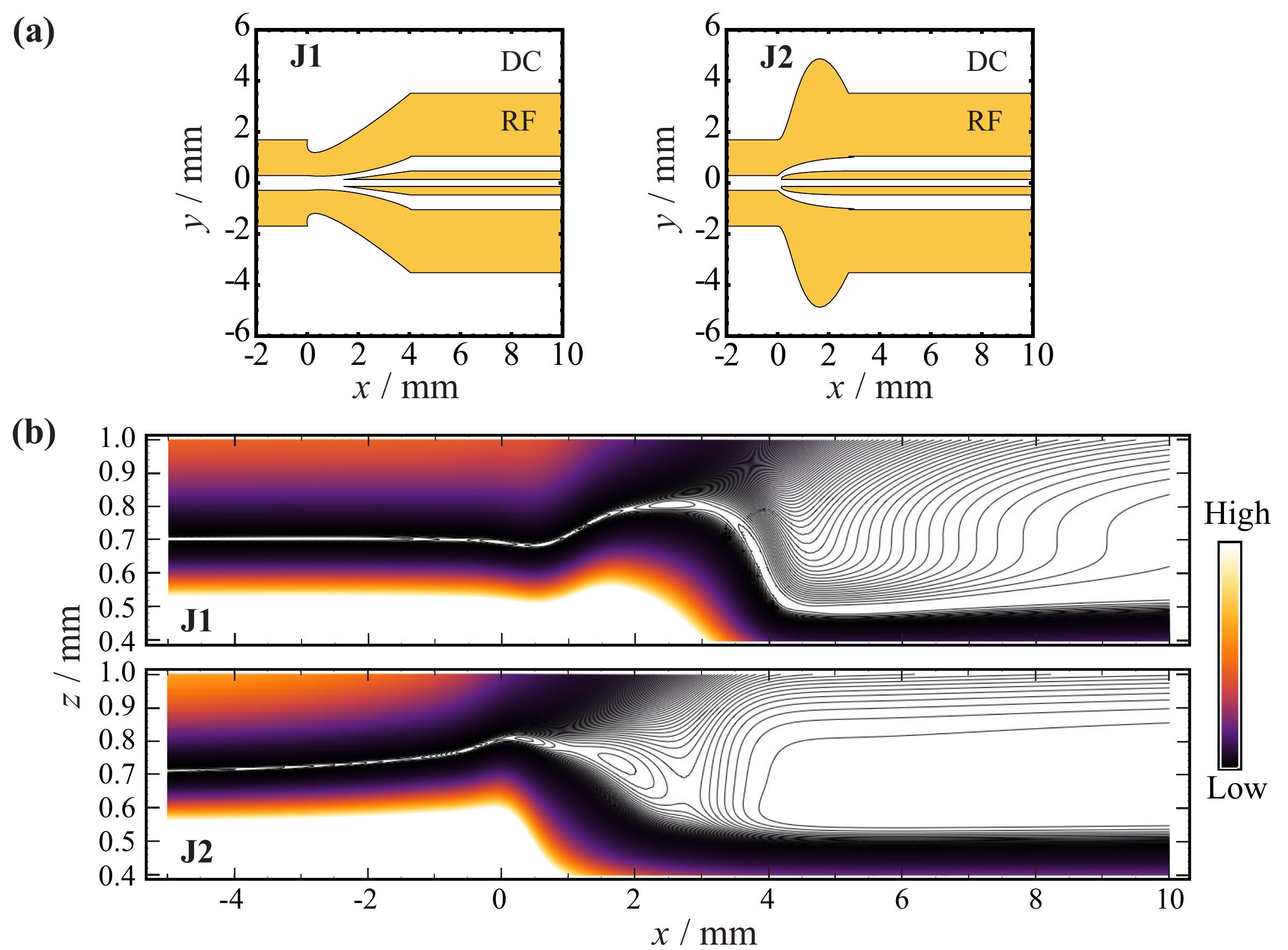}
\centering
\caption{\label{fig:octuplarjunctions-realsize}(a) Optimised octupolar-to-quadrupolar junction geometries (RF (yellow) and DC (white) electrodes). (b) The pseudopotential contour plots for J1 and J2 in the (\textit{x,z}) plane. The junction J2 yields an effectively reduced shuttling path between the quadrupolar and octupolar channels.}
\end{figure}
To design a quadrupole-to-octupole junction that smoothly converts the trapping fields, similar optimisation processes were implemented. 
The geometry was parametrised using three cubic splines (section \ref{subsec:design space}) corresponding to 14 degrees of freedom, see appendix A. Two different solutions were obtained: patterns with gradually enlarging RF electrodes and non-trivial electrode geometries with rather complex patterns, which are labelled with J1 and J2 , respectively, in figure~\ref{fig:octuplarjunctions-realsize}(a). The corresponding pseudopotential profiles of the two junctions are presented in figure~\ref{fig:octuplarjunctions-realsize}(b), calculated for $^{40}$Ca$^{+}$ at the RF amplitude V$_{\rm RF}$=300~V and the RF angular frequency $\Omega_{\rm RF}$=$2\pi\times10.0$~MHz, where the asymptotic quadrupolar channel minimum is at 700~$\mu$m above the electrode plane.
\begin{figure}
\includegraphics[scale=0.61]{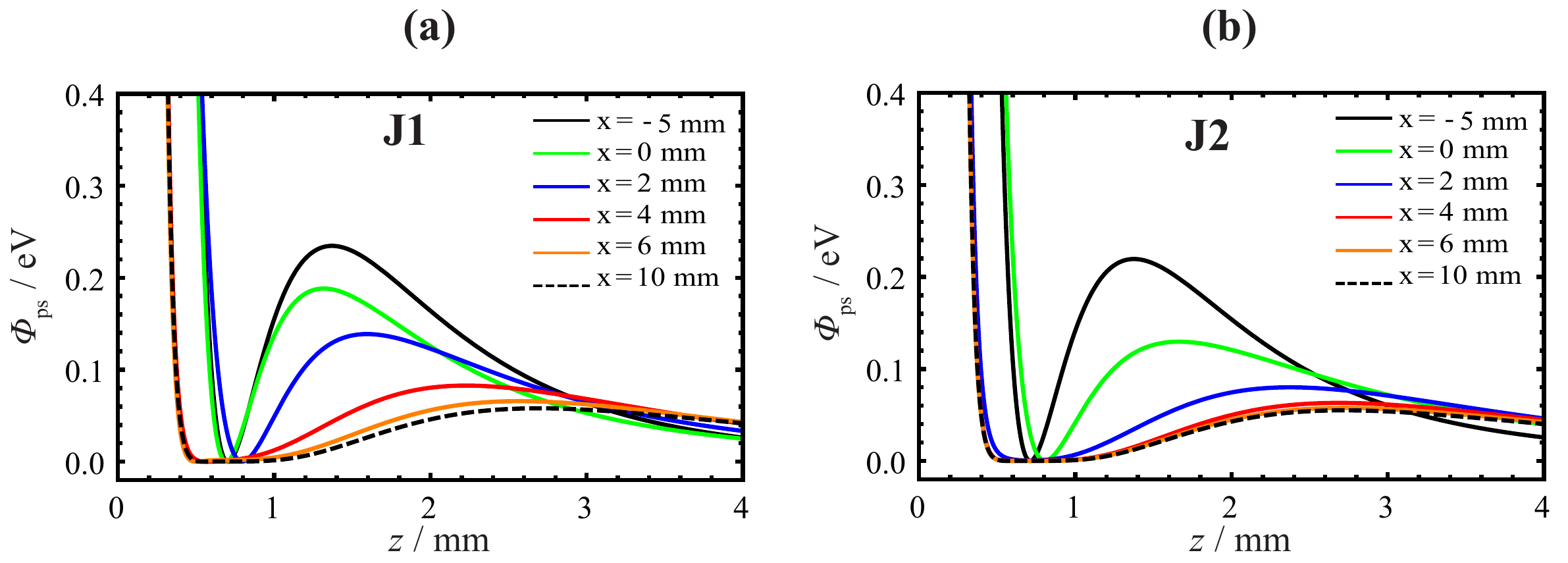}
\centering
\caption{\label{fig:J2andJ4-sixposition} One-dimensional cuts through pseudopotentials for (a) J1 and (b) J2 along the vertical axis \textit{z}, calculated at different locations $x$ (see figure \ref{fig:octuplarjunctions-realsize}) along the ion channel as specified by the legends. For instance, considering the potentials at x=~2~mm, the quadrupolar contribution is more prevalent in J1 than in J2; however, in J2 the octupolar contribution dominates. See text for discussion.}
\end{figure} 
\begin{figure}
\includegraphics[scale=0.57]{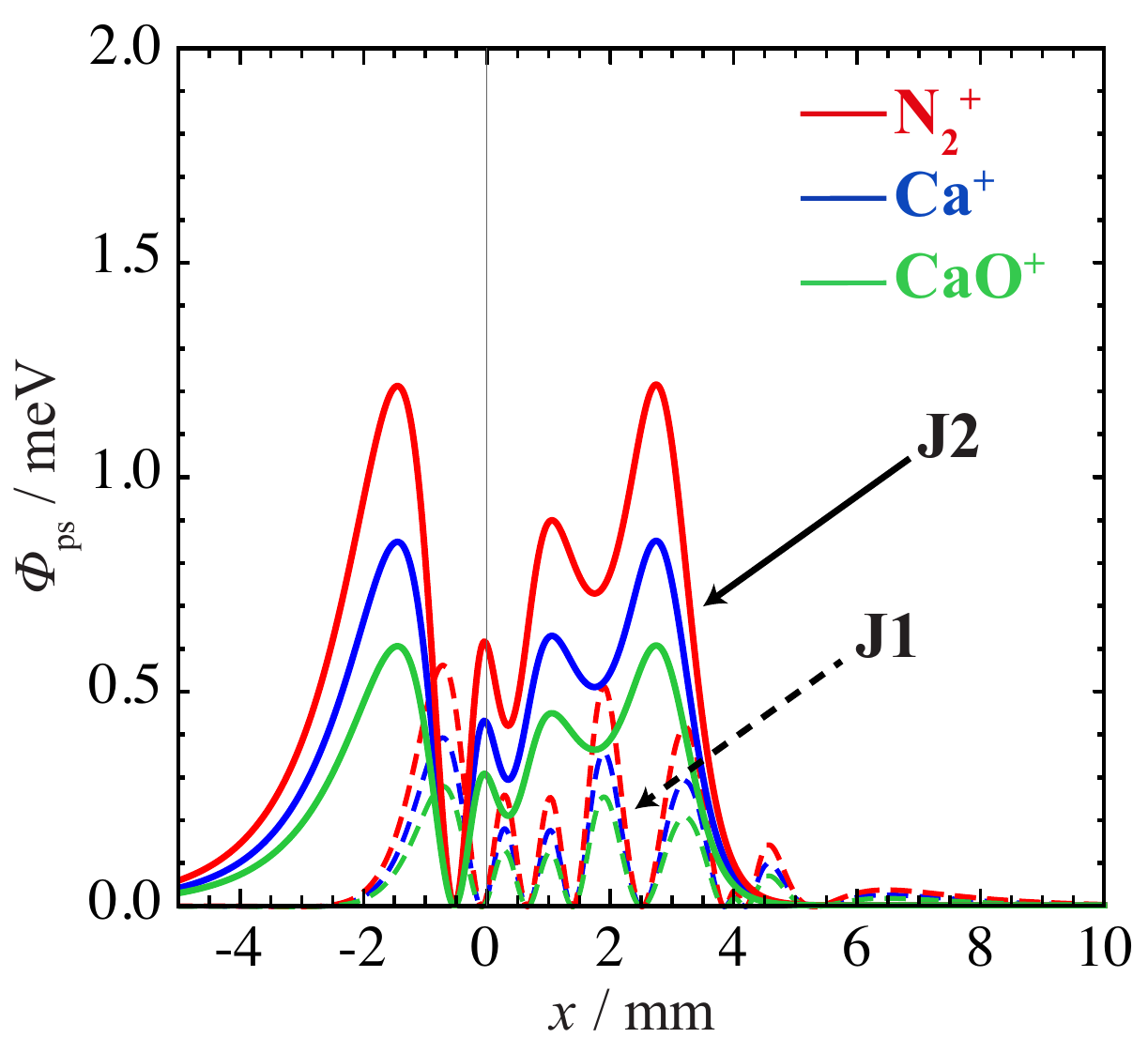}
\centering
\caption{\label{fig:J2andJ4-psbarrier-mass}The pseudopotential barriers along the ion channel for N$_{2}^{+}$, Ca$^{+}$, and CaO$^{+}$ for the optimised junctions J1 (dashed curves) and J2 (solid curves).}
\end{figure} 

J1 was obtained from optimisation processes in which pseudopotential barriers were treated with large weighting factors in equation~(\ref{eqn:obj_total}). Such an electrode pattern can also be intuitively envisioned for a quadrupole-to-octupole junction as it slowly interpolates the two asymptotic channel electrode patterns. In contrast, this transformation occurs on a shorter distance in J2, which was obtained from an optimisation in which the length of the junction was constrained. 
This can be seen in figure~\ref{fig:J2andJ4-sixposition} showing one-dimensional cuts through the pseudopotential along the vertical direction $z$ at six different locations along the ion transport path for J1 and J2.

The pseudopotential barriers calculated along the transport channel for the two structures are compared for N$_{2}^{+}$, Ca$^{+}$, and CaO$^{+}$ in figure~\ref{fig:J2andJ4-psbarrier-mass}. Although the pseudopotential barriers in J2 are larger, the shuttling distance to the octupolar region is shorter making it the preferable geometry. 

The electric fields generated by J2 with the inclusion of 120~$\mu$m-wide gaps were modelled using FEM where only slight refinements were applied in order to avoid sharp electrode shapes. The result ensured a smooth transition between the two trapping regions, and thus no further optimisation was undertaken. Note that in the field-modifying linear junctions the pseudopotential barriers are typically much lower than those of multichannel intersections. 

\section{Design of a multi-functional ion-trap chip for integrated experiments with cold molecular ions}\label{sec:MFIT}
\begin{figure}
\centering
\includegraphics[scale=0.4]{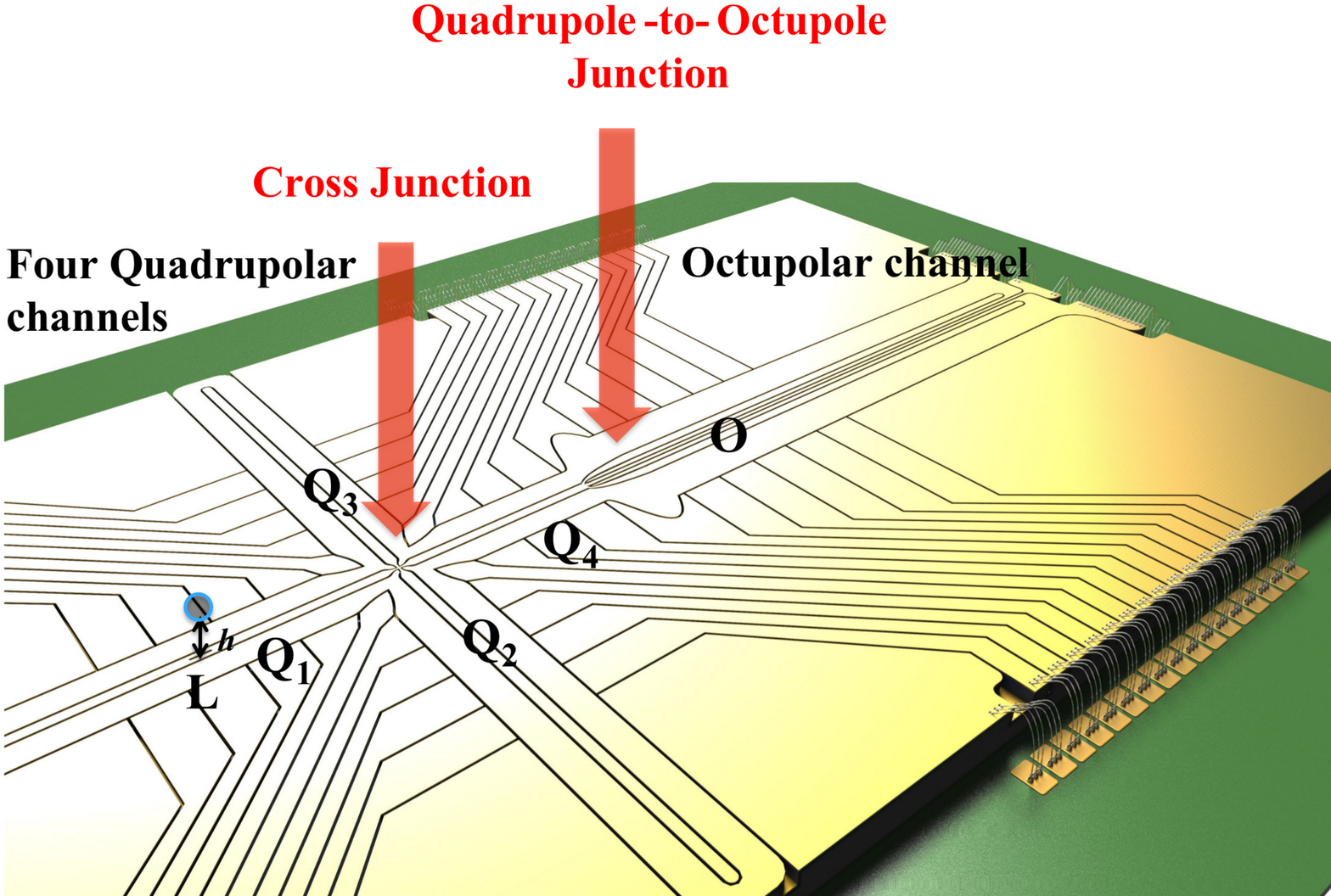}
\caption{\label{fig:mfseit}Overall layout of a multi-functional ion trap chip. L indicates the loading zone where ions are trapped at a height of $h$ above the surface. The four quadrupolar channels are labelled with Q$_{1}$, Q$_{2}$, Q$_{3}$, and Q$_{4}$, and are connected via a cross junction. A quadrupole-to-octupole junction along the fourth quadrupolar arm transforms the trapping field into an octupolar region O.}
\end{figure} 

Building on this study, we have developed a design for a multi-functional SE RF ion-trap chip to enable the performance of various tasks such as loading and preparation of ions, mass spectrometry, spectroscopy, and manipulation of molecular Coulomb crystals on a chip device. 
Figure~\ref{fig:mfseit} illustrates the layout of this chip. 
The chip consists of four quadrupolar ion channels and one octupolar ion channel connected via the two optimised SE junctions discussed in the previous sections. The optimisation of the  SE octupole trap integrated on the chip is discussed in appendix B. 

The proposed trapping architecture features the functionality of a large guided ion-beam apparatus, e.g., see references \cite{asvany08a, armentrout02a}, at a fraction of the production and operation costs. 
The quadrupolar zone Q$_{1}$ can be used to prepare cold molecular ions and to characterise them using, e.g., a resonant-excitation mass spectrometry technique \cite{roth07a}. The experimental sites labelled with Q$_{2}$ and Q$_{3}$ are designed for spectroscopy and ion-neutral reaction studies with suitable accesses for laser and molecular beams. The octupolar region along the fourth arm can be exploited for reaction studies in a large RF-field-free trapping volume as well as for storing ions.
The monolithic, miniaturised design also facilitates the operation of cryogenic systems. 
The concrete experimental implementation of this device will be reported elsewhere.

\section{Conclusions}\label{sec:Conclusion}
We have discussed the design and optimisation of surface-electrode ion-trap junctions of two types, an X-junction between four quadrupoles and a quadrupole-to-octupole junction, in view of developing integrated, monolithic trapping architectures for cold molecular ions. 
A detailed description of the multi-objective optimisation exploited to design arbitrary SE junctions has been presented. The results are modular, scalable components that can be used as design libraries for developing a complex ion-trap network. These optimal geometries are independent of the ion mass and trapping parameters.
We have shown that for a cross junction between four quadrupolar channels, the incorporation of an optimised bridge results in a significant increase of the trap stiffness as well as a better control over the trapping height at the intersection.
Our optimisation yields a non-trivial geometry for the quadrupole-to-octupole junction which features an effectively reduced distance between the two trapping regions.
 
Based on these results, a multi-functional surface-electrode ion-trap chip has been proposed that combines various tasks such as loading and preparation of ions, mass spectrometry, spectroscopy, and manipulation of the structure of crystals in a miniaturised device using precisely shaped trapping potentials. The carefully designed ion channel intersections are intended for the transport of sympathetically cooled molecular ions in the form of bicomponent crystals.

\ack
A.~M. thanks Dr. Bjoern Lekitsch from the University of Sussex for helpful discussions about the microfabrication techniques.
This work has been supported by the University of Basel, the COST Action MP1001 ``Ion Traps for Tomorrow's Applications'', the Swiss National Science Foundation through the National Centre of Competence in Research ``Quantum Science and Technology'' and the Swiss Nanoscience Institute. 

\newpage
\appendix 
\section*{Appendix A: Specifications of the optimised junctions}\label{appendix:obtained parameters}
\setcounter{section}{1}
\begin{figure}
\includegraphics[scale=1.5]{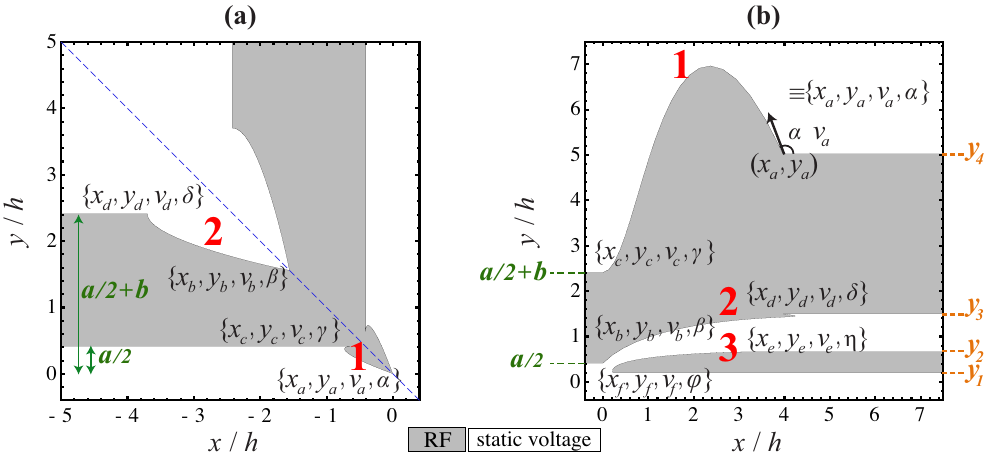}
\centering
\caption{\label{fig:structures}Labelled geometries of the two optimised SE junctions. Due to symmetry, only one quarter of the cross junction and one half of the quadrupole-to-octupole junction are depicted in (a) and (b), respectively. Each cubic spline is specified by the 8 parameters demonstrated in figure~\ref{fig:design-space}(d). These parameters are presented in tables \ref{table:opti_structures_cross} and \ref{table:opti_structures_octu}. $a$ and $b$ are the RF electrode separation and width for the optimal quadrupole, respectively. $y_{1-4}$ are the position of the RF-electrode edges for the optimised SE octupole guide, see appendix B. The dashed line in (a) shows the line of equality in the ($x,y$) plane.}
\end{figure}

\begin{table}
\caption{\label{table:opti_structures_cross}Specification of the optimised cross junction (figure \ref{fig:crossjunction-realsize}(a)), see the labelled cubic splines in figure~\ref{fig:structures}(a). $a$ and $b$ are the RF electrode separation and width for the optimal SE quadrupole, respectively (table \ref{table:opti_quadchanel}). Note that $\lbrace x_{a}, y_{a}\rbrace$ and $\lbrace x_{b}, y_{b}\rbrace$ were constrained to remain on the line of equality (i.e., $x=y$), and thus the geometry is specified by 12 degrees of freedom. The length scales are in units of the trapping hight $h$, and the unit of angles is radian.}
\renewcommand{\arraystretch}{1.0}
\begin{indented}
\item[]\begin{tabular}{@{}*{12}{@{\extracolsep{0pt plus16pt}}l}}
\hline
\mr 
Cubic spline 1 \qquad & $\lbrace x_{a}, y_{a}, v_{a}, \alpha \rbrace$ & $\lbrace x_{c}, y_{c}, v_{c}, \gamma \rbrace$ \\ 
&$\lbrace 0.004,0.004, 2.242,0.353\rbrace$ & $\lbrace 0.643,$ fixed at $a/2, 0.633,3.091 \rbrace$ \\ \mr
Cubic spline 2 \qquad & $\lbrace x_{b}, y_{b}, v_{b}, \beta \rbrace$ & $\lbrace x_{d}, y_{d}, v_{d}, \delta \rbrace$\\  
& $\lbrace 1.562, 1.562, 0.064,-0.172 \rbrace$ & $\lbrace3.690, $ fixed at $a/2+b, 1.25098, 1.637 \rbrace$\\ \br \hline
\end{tabular}
\end{indented}
\end{table}
\begin{table}
\caption{\label{table:opti_structures_octu}Specification of the quadrupole-to-octupole junction, see figure~\ref{fig:structures}(b). $y_{i}$ are the RF-electrode edges positions obtained from the optimisation of the SE octupole discussed in appendix B, and $a$ and $b$ are the RF electrode separation and width for the quadrupole guide, respectively. Note that due to the constraints used, e.g., $x_{a}=x_{d}=x_{e}$, the geometry was specified by 14 degrees of freedom. The length scales are in units of the trapping hight $h$, and the unit of angles is radian.}
\renewcommand{\arraystretch}{1.0}
\begin{indented}
\item[]\begin{tabular}{@{}*{12}{@{\extracolsep{0pt plus16pt}}l}}
\hline
\mr
\lineup
Cubic spline 1 \quad & $\lbrace x_{a}, y_{a}, v_{a}, \alpha \rbrace$ \quad & $\lbrace x_{c}, y_{c}, v_{c}, \gamma \rbrace$ \\ 
&$\lbrace 3.969,$ fixed at $y_{4}, 18.420, 2.008 \rbrace$ \quad & $\lbrace$fixed  at $0, $ fixed at $a/2+b, 2.539, 3.141 \rbrace$ \\ \mr
Cubic spline 2 \quad & $\lbrace x_{b}, y_{b}, v_{b}, \beta \rbrace$ & $\lbrace x_{d}, y_{d}, v_{d}, \delta \rbrace$\\  
& $\lbrace $fixed  at $0, $ fixed  at $a/2, 5.204, 0.794 \rbrace$ & $\lbrace 3.969, $ fixed  at $y_{3}, 4.107, 3.024 \rbrace$\\ \mr 
Cubic spline 3 \quad & $\lbrace x_{e}, y_{e}, v_{e}, \eta \rbrace$ & $\lbrace x_{f}, y_{f}, v_{f}, \phi \rbrace$\\  
& $\lbrace 3.969, $ fixed  at $y_{2}, 0.0, 1.452 \rbrace$ & $\lbrace 0.215,$ fixed  at $ y_{1}, 1.634, -1.127 \rbrace$\\ \br\hline 
\end{tabular}
\end{indented}
\end{table}

The electrode geometries of the optimised cross junction (figure \ref{fig:crossjunction-realsize}(a)) and quadrupole-to-octupole junction J2 (figure \ref{fig:octuplarjunctions-realsize}(a)) obtained are fully defined by the cubic splines specified in tables \ref{table:opti_structures_cross} and \ref{table:opti_structures_octu}, respectively. For clarity, each spline is labelled in figure~\ref{fig:structures}, see also figure~\ref{fig:design-space}(d) for the definition of the parameters used. Note that the dimensions were normalised, e.g., lengths are in units of the trapping height, and therefore, the values can be accordingly scaled to suit any required size of junction  architectures.

For the optimisation of the cross junction, the transformation coefficients in equation~(\ref{eqn:obj_total}) were calculated to be $\lbrace \xi_{1}$,  $\xi_{2}$, $\xi_{3}\rbrace$~=~$\lbrace$1.6$\times10^{-6}$, 6.1$\times10^{-12}$, 1.0$\times10^{-4}\rbrace$ associated with pseudopotential barriers and gradients, and the trapping height variation, respectively. 
The corresponding weighting factors were altered in the interval $(0,1)$ such that the three objective functions were significantly represented without deterministic priority, e.g., in this case $\lbrace w_{1}$,  $w_{2}$, $w_{3}\rbrace$~=~$\lbrace 0.2$,  $0.6$, $0.2\rbrace$. For the quadrupole-to-octupole junction, these values were given by $\lbrace \xi_{1}$,  $\xi_{2}$, $\xi_{3}\rbrace$~=~$\lbrace$2.2$\times10^{-5}$, 9.9$\times10^{-11}$, 8.9$\times10^{-4}\rbrace$ and $\lbrace w_{1}$,  $w_{2}$, $w_{3}\rbrace$~=~$\lbrace 0.375$,  $0.5$, $0.125\rbrace$.

\appendix
\section*{Appendix B: Optimisation of a SE octupole channel}
\label{appendix:octupole second}
\begin{figure}
\includegraphics[scale=0.18]{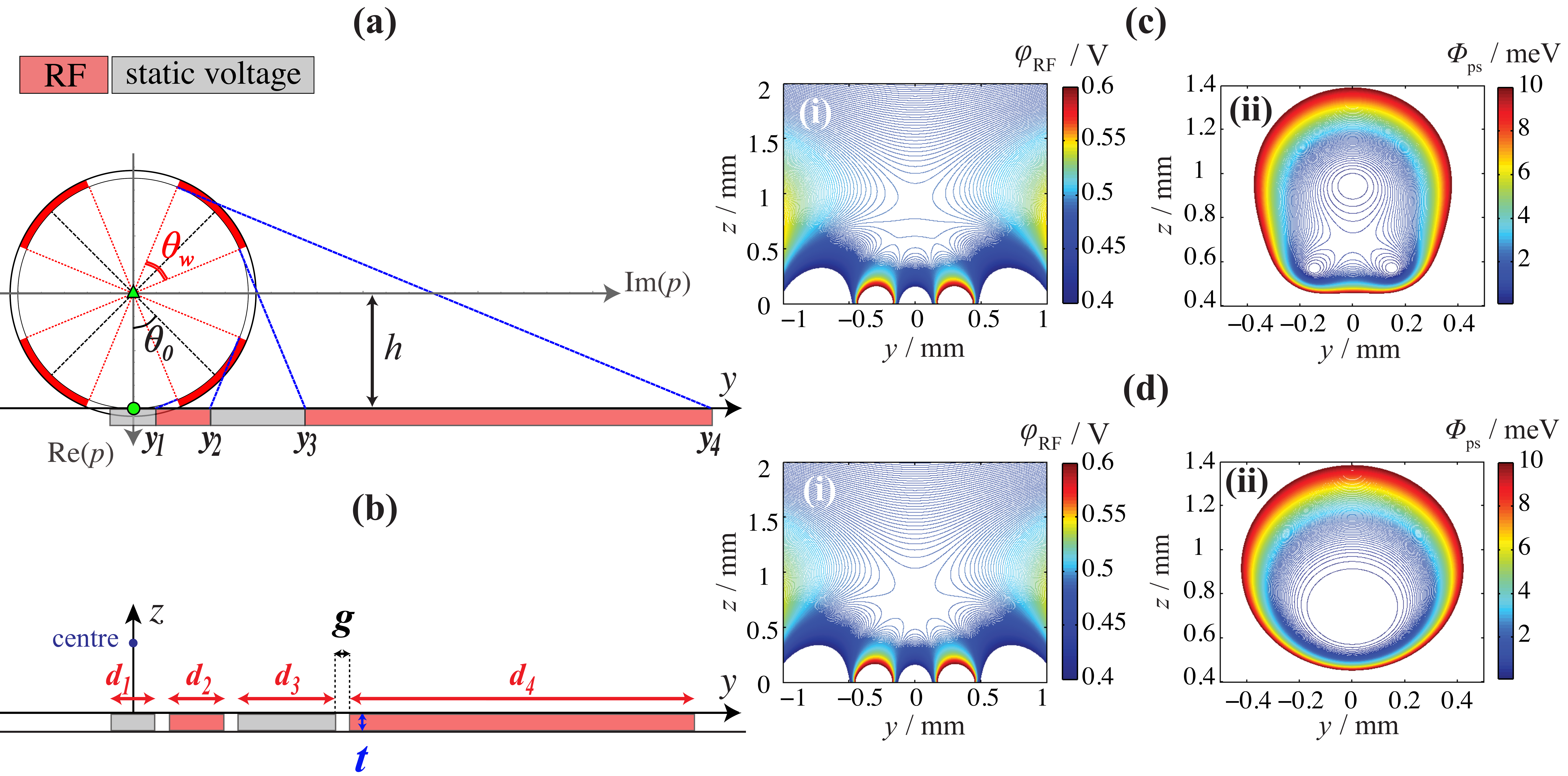}
\centering
\caption{\label{fig:wesenbergmodel}(a) Graphical representation of conformal mapping of an octupole onto a surface. The green triangle and the circle indicate the trap centre and the point on the electrode plane right below the trap centre, respectively, and $h$ is the trapping height. The configuration illustrates the optimal octupole of maximum strength for the analytical solution $\theta_{w}$=$\pi/4$ and $\theta_{0}$=$\pi/4$. The electrode edge positions $\lbrace y_{1}, y_{2}, y_{3}, y_{4} \rbrace$ are given by the circle's tangents extended to the $y$ axis. Due to the mirror symmetry, negative $y$ values are not shown. (b) A schematic of the electrode configuration for a SE octupole trap showing the parameters used in the numerical optimisation. (c) and (d) Results of the finite elements analysis for the two optimal geometries obtained from analytical and numerical optimisation, respectively: (i) the RF basis function in the transverse plane, and (ii) the pseudopotential for $^{40}$Ca$^{+}$ in which contours are separated by 0.1 meV.}
\end{figure}

\setcounter{section}{2} 
Three approaches can be considered to find the optimal geometry of a SE multipole trap within the "gapless plane approximation" \cite{schmied10a}. 
In the first approach, the \textit{OptimalFinitePattern} routine from the \textit{SurfacePattern} package enables the computation of an optimal surface geometry for a set of given local constraints, i.e., fixed values of the potential and derivatives at a certain point based on the relaxed Linear Programming method \cite{LinearProgramming97}.
In the second approach, the optimum position of the electrode edges is analytically calculated. We maximised the strength of the RF-electrodes basis function $\mathit{\Theta}_{\rm RF}$ at a given height $h$ while the lower order multipole coefficients were set to zero:

\begin{equation} 
\label{eqn:OCTU-analytical}
\fl
\cases{
\eqalign{ 
{\rm \maximise \limits_{y_{1}, \ldots , y_{p}}}\quad \frac{\partial^{n}\mathit{\Theta}_{\rm RF}(y_{1}, ..., y_{p} )}{\partial z^{n}} \\
{\rm subject \ to}\quad \frac{\partial^{i+j+k}\mathit{\Theta}_{\rm RF}(y_{1}, ..., y_{p} )}{\partial x^{i}\partial y^{j}\partial z^{k}}=0,}}
\end{equation}
where $\lbrace i, j,k \rbrace\in \lbrace \mathbb{N} ~\vert ~\exists n: (i+j+k) < n \rbrace$ and $n$ denotes the order of the multipole and $p$ is the number of degrees of freedom. For an optimal octupole ($n=4$) with symmetric design of four RF electrodes ($p=4$), the solution of equation~(\ref{eqn:OCTU-analytical}) is given by $\lbrace y_{1}, y_{2}, y_{3}, y_{4} \rbrace$~=~$h\times\lbrace -1-\sqrt{2}+\sqrt{2 \left(2+\sqrt{2}\right)},1-\sqrt{2}+\sqrt{4-2 \sqrt{2}},-1+\sqrt{2}+\sqrt{4-2 \sqrt{2}},1+\sqrt{2}+\sqrt{2 \left(2+\sqrt{2}\right)} \rbrace $.
The third approach is based on conformal mapping of conventional cylindrical trap geometries to strip electrodes on a plane \cite{wesenberg08a}. 
Figure~\ref{fig:wesenbergmodel}(a) depicts the configuration of an optimal octupole trap featuring maximum absolute value of $\partial^{4} \mathit{\Theta}_{\rm RF} / \partial z^{4}$. 
All three approaches generate the same result \cite{mokhberi16a}. 

The geometry obtained was analysed using FEM with the inclusion of the realistic inter-electrode spacings. The result showed a non-vanishing RF field at the trapping centre which yield three pseudopotential minima (figure~\ref{fig:wesenbergmodel}(c)). 
The ramification would be disordered structures of Coulomb crystals with ions driven to unwanted regions, those in which atomic ions are not efficiently cooled by lasers. 
Attempts to compensate this effect by the application of static fields would be inadequate because it is solely caused by incorporating gaps to the geometry.

To address this issue, electric potentials obtained from FEM calculations were used to generate numerical objective functions for a multi-objective optimisation using the Nelder-Mead algorithm. 
The merit function was given by:
\begin{eqnarray}
F (d_{1}, d_{2}, d_{3}, d_{4})=\lambda_{1} f ^{(1)}+\lambda_{2} f ^{(2)}+\lambda_{3} f ^{(3)}-\lambda_{4} f ^{(4)},
\end{eqnarray}
where the objective functions $f^{(n)}$ were formulated as \(f^{(n)}=\frac{\partial^{n}\mathit{\Theta}_{\rm RF} (d_{1}, d_{2}, d_{3}, d_{4})}{{\partial z}^n}\). The parameters $d_{1}, d_{2}, d_{3}$, and $d_{4}$ are schematically shown in figure~\ref{fig:wesenbergmodel} (b). 
$\lambda_{i}$ stands for the weighting factor including a transformation coefficient, see section~\ref{subsec:objfunc}. The result of the second optimisation  is presented in figure~\ref{fig:wesenbergmodel} (d).
As the gap aspect ratio $\frac{t}{g}$ decreases particularly for $\frac{t}{g}<2$, the residual RF field is enhanced \cite{mokhberi16a}, where $t$ and $g$ are the inter-electrode spacing and the trench depth, respectively, see figure~\ref{fig:wesenbergmodel}(b). 
The explanation is provided by understanding the role of the gap polarization and interpolation potentials that alter the surface potential \cite{schmied10a}. 
The effect is pronounced in the octupolar channel because of the sensitivity of the field-free region to the variation of the surface potential.

\section*{References}
\providecommand{\newblock}{}

\end{document}